\documentclass[runningheads]{llncs}

\usepackage[nocompress]{cite}
\usepackage{booktabs} 
\usepackage{xspace}
\usepackage{balance}
\usepackage{tablefootnote}
\usepackage{comment}

\usepackage{pgfplots}
\pgfplotsset{compat=1.9}

\usepackage{listings}
\usepackage{subfig}

\usepackage[shortcuts]{extdash}

\usepackage{tikz}
\newcommand*\circledb[1]{\tikz[baseline=(char.base)]{ 
    \node[shape=circle,draw,inner sep=.5pt] (char) {\textcolor{black}{\scriptsize #1}};}}

\newcommand{\code}[1]{\texttt{\small#1}}
\newcommand{\tperc}[1]{$#1$\,\%} 
\newcommand{\perc}[1]{$#1$\%}

\newcommand{\toolname}{PartiSan}

\usepackage{xspace}
\usepackage{relsize}
\def\ifmonospace{\ifdim\fontdimen3\font=0pt }  
\def\cpp{%
\ifmonospace%
    \C++%
\else%
    C\kern-.1167em\raise.30ex\hbox{\smaller{++}}%
\fi%
\spacefactor1000 }

\newcommand{\arch}{x86\=/64}
\newcommand{\asan}{\mbox{ASan}}

\newcommand{\ubsan}{\mbox{UBSan}}
\newcommand{\spec}{\mbox{SPEC}}
\newcommand{\speclong}{SPEC CPU\,2006}
\newcommand{\libfuzzer}{libFuzzer}  

\begin{document}
\title{\toolname: Fast and Flexible Sanitization via Run-time Partitioning}
\titlerunning{\toolname: Fast and Flexible Sanitization via Run-time Partitioning}
\author{Julian Lettner \and Dokyung Song \and Taemin Park \and Stijn Volckaert \and Per Larsen \and Michael Franz}
\institute{University of California, Irvine\\
\email{\{julian.lettner, dokyung.song, tmpark, stijnv, perl, franz\}@uci.edu}}
\authorrunning{Lettner et al.}

\maketitle

\begin{abstract}
Sanitizers can detect security vulnerabilities in C/\cpp{} code that elude static analysis.
Current practice is to continuously fuzz and sanitize internal pre-release builds. Sanitization-enabled builds are rarely released publicly. This is in large part due to the high memory and processing requirements of sanitizers.

We present \toolname{}, a run-time partitioning technique that speeds up sanitizers and allows them to be used in a more flexible manner.
Our core idea is to partition the execution into sanitized slices that incur a run-time overhead, and ``unsanitized'' slices running at full speed.
With \toolname{}, sanitization is no longer an all-or-nothing proposition. A single build can be distributed to every user regardless of their willingness to enable sanitization and the capabilities of their host system.
\toolname{} can automatically adjust the amount of sanitization to fit within a performance budget or disable sanitization if the host lacks sufficient resources.
The flexibility afforded by run-time partitioning also means that we can alternate between different types of sanitizers dynamically; today, developers have to pick a single type of sanitizer ahead of time.
Finally, we show that run-time partitioning can speed up fuzzing by running the sanitized partition only when the fuzzer discovers an input that causes a crash or uncovers new execution paths.

\end{abstract}

\section{Introduction}

Although modern, safe languages could gradually replace C/\cpp, the sheer amount of legacy systems code forces security researchers to search for and fix memory corruption vulnerabilities in existing code in the near term.
While some bugs can be found through static program analysis, many cannot.
Sanitizers are dynamic analysis tools that can detect memory corruption and many other problems as well as pinpoint their occurrence during program execution~\cite{ASAN,UBSAN}.
To increase coverage, sanitizer runs can be driven by a fuzzer. A fuzzer simply feeds the program random inputs and records inputs that generate crashes or cause previously unexecuted code to run.

Sanitizers instrument programs---usually during compilation---to detect issues such as memory corruption and undefined behavior.
This instrumentation incurs significant overheads, so sanitizers are turned off in release builds and traditionally only enabled on internal quality assurance builds that run on high-end hardware.
This is less than ideal as the number of paths executed by test suites and fuzzers is outnumbered by the number of paths executed by end users.

In a recent experiment, the Tor Project released sanitizer-enabled (labeled ``hardened'') builds directly to its users~\cite{koppen2017hardenedTB}.
The hardened build series was discontinued in part due to the high performance overhead and in part due to confusion among end users over which version to download.
With access to \toolname{}, the Tor Project developers could have released builds that automatically adapt the level of sanitization to the capabilities of the host system.
Overhead can be limited by using a conservatively low, adaptive threshold by default (and possibly disabling sanitization completely on underpowered systems)
while simultaneously allowing expert users to modify the default settings (thereby also eliminating the need for multiple build versions).

\toolname{} clones frequently executed functions at compile time and efficiently switches among them at run time.
Each function variant can be optimized and sanitized independently, and thus has different security and performance properties.
In the simplest case, one variant is instrumented to sanitize memory accesses while the other one is not.
\toolname{} supports configurable run-time partitioning policies that determine which variant is invoked when a function is called.
For example, \toolname{} can execute slow variants (e.g., variants with expensive checks) with low probability on frequently executed code paths, and with high probability on rarely executed paths.
This policy helps us keep the sanitization overhead below a given threshold.

This is superficially similar to the ASAP framework by Wagner et al.~\cite{ASAP} insofar that both approaches explore the idea of reducing the amount of sanitization on the hot path. However, ASAP \emph{statically} partitions the code into parts with or without sanitization based on previous profiling runs at compile time. \toolname\ prepares programs for partitioning at compile time but does the partitioning \emph{dynamically} at run time. This allows us to produce a single binary that adapts to each individual host system, sanitizing as many paths as possible under a given performance budget. Moreover, we can create $N$ different function variants to support $N-1$ types of sanitization in a single binary. Table~\ref{tab:asap_concept} contrasts \toolname{} and ASAP\@.
Both our work and ASAP build on the assumption that security vulnerabilities in frequently executed code get discovered and patched relatively quickly, whereas vulnerabilities in rarely executed code might go unpatched for a long time.

\begin{table*}
  \centering
  \caption{Conceptual comparison of ASAP and \toolname}\label{tab:asap_concept}
  \begin{tabular}{lll}
    \toprule

      Statement                           & ASAP                                  & \toolname{} \\  
    \midrule
      Goal                  \ldots  & deploy sanitizers as mitigations\hspace{5pt} & find bugs efficiently  \\ 
      Partitioning is                 \ldots  & static (compile time)             & dynamic (run time) \\ 
      Overhead reduction      \ldots  & removal of expensive checks   & probabilistic checking \\ 
      Code is                         \ldots  & deleted                           & cloned \\ 
      Assertions are       \ldots  & removed                           & retained \\ 
      Detect bugs in cold code    \ldots  & always                            & always            \\ 
      Detect bugs in hot code     \ldots  & never                             & probabilistically \\ 
    \bottomrule
  \end{tabular}
\end{table*}

This paper makes the following contributions:
\begin{itemize}
\item We describe \toolname{}\footnote{\toolname{} will be made available upon acceptance of this paper.}\hspace{-3pt}, a framework to partition program execution into sanitized/unsanitized fragments at run time.
  Unlike previous approaches, the partitioning is not static but happens dynamically according to a policy-driven, run-time partitioning mechanism which selects the function variant to execute with low overhead.
  This lets developers release sanitizer-enabled builds to end users and thereby cover more execution paths.

\vspace{.5em}\item We present a fully-fledged prototype implementation of our ideas and explore three concrete run-time partitioning policies.
  We combined \toolname{} with two sanitizers and measured the performance overhead on the \speclong{} benchmark suite with our expected-cost partitioning policy.

\vspace{.5em}\item We present a thorough evaluation showing that our approach still detects the majority of vulnerabilities at greatly reduced performance overheads.
  For the popular \asan\ and \ubsan\ sanitizers, \toolname\ reduces overheads by \perc{68} and \perc{76} respectively.

\vspace{.5em}\item We demonstrate an important use case of \toolname{}: improving fuzzing efficiency.
  We combined \toolname{} with a popular fuzzer and measured consistently increased fuzzing throughput.
\end{itemize}

\section{Background}\label{sec:background}

LLVM~\cite{LLVM}, the premier open-source compiler, includes five different sanitizers.
We demonstrate \toolname{} by applying two of these sanitizers to a variety of programs.
\asan, short for AddressSanitizer~\cite{ASAN}, instruments memory accesses and allocation operations to detect a range of memory errors, including spatial memory errors such as out-of-bounds accesses and temporal violations such as use-after-free bugs.
\ubsan, short for UndefinedBehaviorSanitizer~\cite{UBSAN}, currently detects $22$ types of operations whose semantics are undefined~\cite{UBinLLVM} by the C standard~\cite{ISOC}.
\ubsan\ includes checks for integer overflows, uses of uninitialized or unaligned pointers, and undefined integer shifts.

We used these two sanitizers with \toolname{} for two reasons.
First, the combination of \asan\ and \ubsan\ detects many of the vulnerabilities that are security critical.
Second, both sanitizers can be applied selectively.
Removing any of the sanitization checks from a program does not affect the correct functioning of the remaining checks.
This makes these sanitizers a good fit for our framework, in which we selectively skip sanitization through run-time partitioning.

\section{Design}

\begin{figure*}[t!!]
  \centering
  \includegraphics[width=1.0\textwidth]{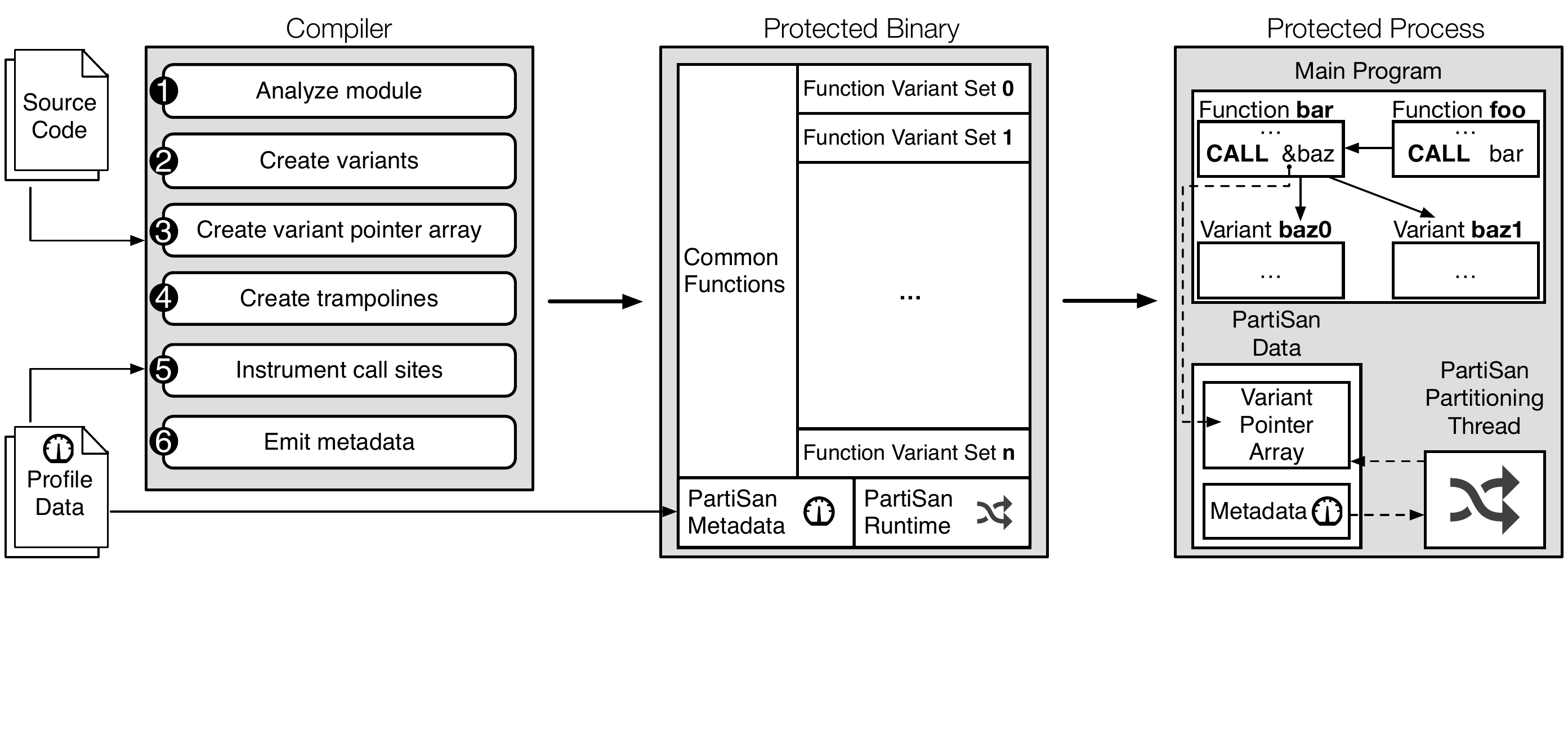}
  \vspace{-6.0em}
  \caption{System overview. The compiler (left) creates \toolname-enabled applications (center) that have multiple variants of each function.
    A run-time indirection through the variant pointer array (right) ensures that the control flow calls the currently active variant.
    \toolname's runtime periodically activates function variants according to the configured partitioning policy.}\label{fig:overview}
\end{figure*}

Our goal is to reduce the run-time overhead of the sanitizers.
We do this by creating multiple variants of each function, applying sanitizers to some variants, and embedding a runtime component that partitions the execution of the program into sanitized/unsanitized slices based on a policy.

Figure~\ref{fig:overview} shows an overview of the \toolname{} system.
To apply \toolname\ to an application, the developer must compile the source code of the program with our modified compiler (left side of Fig.\,\ref{fig:overview}).
Some partitioning policies require that the developer supply profile data.

The compiler generates an application with multiple variants for each function.
To simplify the following discussion, we will focus on use cases where we generate two variants.
One of the variants, which we refer to as the \emph{unsanitized variant}, does not include any sanitizer checks. 
The other variant, which we call the \emph{sanitized variant}, incorporates all sanitizer instrumentation.

The compiler modifies the program's control flow as follows.
Rather than calling functions directly, the functions call each other through an additional level of indirection.
Specifically, the compiler embeds a ``variant pointer array'' containing one slot for each function in the program source code.
At run time, each slot holds the pointer to the currently active variant of the corresponding function.
The \toolname{} runtime, which is linked into the application by our compiler, selects and activates one variant of each function according to the configured partitioning policy.

The runtime currently supports three partitioning policies: random partitioning, profile-guided partitioning, and expected-cost partitioning.
With the random partitioning policy, the runtime randomly selects the active variants, whereas
the profile-guided and expected-cost partitioning policies select active variants with a probability that depends on the execution frequency (``hotness'') and/or expected sanitization cost of that function.
These policies can help us limit the cost of sanitization.

\subsection{Creating Function Variants}

\toolname's compiler pass runs after the source code is parsed and converted into intermediate representation (IR) code.
As its first step (Step \circledb{1} in Figure~\ref{fig:overview}), our compiler pass analyzes the IR code and determines which functions to create variants for.
We do not necessarily create multiple variants for each function.
If the developer selects the profile-guided or expected-cost partitioning policy, and if the profile data indicates that a function is infrequently executed, then we create only the sanitized variant for that function.
This design choice prevents \toolname{} from unnecessarily inflating the code size of the program and is justified because checks in infrequently executed code have little impact on the program's overall performance.

Then, \toolname{} creates the function variants (\circledb{2}).
First, we clone functions that should have two variants and give them new, unique names.
Then, we apply the requested instrumentations to the variants.

\subsection{Creating the Indirection Layer}

Once the function variants are created, our compiler pass creates the indirection layer, through which we route all of the program's function calls.
This ensures that the program can only call the active variant of each function.
Our indirection layer consists of three components: the variant pointer array (right side of Figure~\ref{fig:overview}), trampolines, and control-flow instructions that read their target from the pointer array.

Our compiler starts by embedding the variant pointer array into the application (\circledb{3}).
The pointer array contains one slot for each function that has multiple variants.
Each slot contains a pointer to the entry point of the currently active variant of that function.

Then, we create trampolines for externally reachable and address-taken functions (\circledb{4}).
A trampoline jumps to the currently active variant of its associated function.
We assign the original name of the associated function to the trampoline.
This way, we ensure that any call that targets the original function now calls the trampoline, and consequently, the currently active variant of the original function instead.

Finally, we transform all direct call instructions that target functions with multiple variants into indirect control-flow instructions that read the pointer to the active variant of the target function from the pointer array (\circledb{5}).
This optimization eliminates the need to route direct calls within the program through the function trampolines.
However, the trampolines may still be called through indirect call instructions, or by external code.

\subsection{Embedding Metadata}

Our compiler embeds read-only metadata describing each function and its variants into the application (\circledb{6}).
The metadata can consist of the function execution frequencies read from the profile data, the estimated execution costs for all function variants, and information connecting each slot in the variant pointer array to the variant entry points associated with that slot.
Our partitioning mechanism bases run-time decisions on the metadata.

\subsection{The \toolname{} Runtime}\label{sec:policy}

Our runtime implements the selected partitioning policy by activating and deactivating variants.
While a specific variant is active, none of the other variants of that same function can be called.
To activate a variant, our runtime writes a pointer to that variant's entry point into the appropriate slot in the pointer array.
\toolname{} periodically activates variants on a background thread.
This allows us to implement a variety of partitioning policies that do not slow down the application thread(s). 
Operating on a background thread also allows our runtime to run frequently, and thus make fine-grained partitioning decisions.

\paragraph{\textbf{Random Partitioning}}
With this policy, our runtime component activates a randomly selected variant of each function whenever our thread wakes up.
Since we only generate two variants of each function, this policy divides the execution time evenly among the sanitized/unsanitized function variants.

\paragraph{\textbf{Profile-Guided Partitioning}}
With this policy, our runtime component collects the list of functions with multiple variants in the program and orders this list based on the functions' execution counts recorded during profiling.
Our runtime activates the sanitized variant of a function with a probability that is inversely proportional to its order in the execution count list.
The sanitized variant of the most frequently executed function is activated with 1\% probability, and that of the least frequently executed function with a 100\% probability.
Note that this partitioning policy does not estimate the overhead impact of executing a sanitized variant instead of an unsanitized variant.
It also does not consider the absolute execution count of a function.
For example, the second least executed function in a program with $100$ functions is sanitized 99\% of the time, even if its execution count is $1000$ times higher than that of the least executed function.

\paragraph{\textbf{Expected-Cost Partitioning}}
This policy improves upon the profile-guided partitioning policy by calculating sanitization probabilities based on function execution counts (read from the profile data) and estimated sanitization cost.
We estimate the cost of sanitization for each function by calculating the costs of all function variants using LLVM's Cost Model Analysis. 
We then calculate the probability of activating the sanitized variant for a function using formula:
$$P_{sanitization}(f) = \frac{sanitization\:budget(f)}{cost_{sanitization}(f) * execution\:count(f)}$$

The sanitization overhead budget is chosen by the developer and is evenly distributed among the functions in the program.

\section{Implementation}
Our prototype implementation of \toolname{} supports applications compiled with clang/LLVM\,5.0~\cite{LLVM} on the \arch{}~architecture.
Our design, however, is fully generalizable to other compilers and architectures.

\subsection{Profiling}

Two of our run-time partitioning policies rely on profile data to calculate the sanitization probabilities.
We use LLVM's built-in profiling functionality to generate binaries that collect profile data.

\subsection{Compiler Pass}

Our pass instruments the program code at the LLVM IR level processing one translation unit at a time.
\toolname{} is fully compatible with standard build systems and program loaders.
We scheduled our pass to run right before the LLVM~sanitizer passes, which run late in the compiler pipeline.
This allows us to define (mostly declaratively) which variants get instrumented without interfering with LLVM's earlier optimization stages.

\paragraph{\textbf{Creating Function Variants}}
Of the sanitizers bundled with LLVM, our pass currently supports \asan\ and \ubsan.
We did not modify any sanitizer code and most of \toolname's code is tool-agnostic.
To create the function variants, we begin by passing the necessary \code{-fsanitize} command line options to the compiler.
\asan's front-end pass prepares the program by marking all functions that require sanitization with a function-level attribute.
With just one line of \asan-specific code, \toolname{} removes this function attribute for the unsanitized variants.
\ubsan's front-end pass embeds many of its checks before the program is translated into IR\@.
\toolname{} contains $56$ lines of code to remove these checks from the unsanitized variants.

\paragraph{\textbf{Creating the Indirection Layer}}
We create the indirection layer as follows.
We begin by collecting the set of functions that have multiple variants.
Then, we add the variant pointer array as a global variable with internal linkage.
We choose the size of the array such that it has one slot for every function in the set.
Next, we create trampolines for all functions in the set.
The trampoline, which takes over the name of the function it corresponds to, forwards control to the currently active variant of that function.
By taking over the name of the original function, the trampoline ensures that any calls to that function will be routed to the currently active variant.

Next, we replace all call instructions that target functions in the set with indirect call instructions that read their call target from the variant pointer array.
Functions outside the compilation unit will not be in the set, but might still have multiple variants.
While we do not replace calls to such instructions, the call will still (correctly) invoke the currently active variant of the target function because it will be routed through that function's trampoline.

Note that the compiled program will only contain the trampolines that may actually be used at run time.
If a trampoline's corresponding function is not externally visible (and thus cannot be called by external code) and it does not have its address taken (and thus cannot be called indirectly), then the trampoline will be deleted by LLVM's dead code elimination pass.

Figure~\ref{fig:machine_code} shows the assembly code that is generated for the trampolines and transformed call sites.

\newsavebox{\origcallbox}
\begin{lrbox}{\origcallbox}
  \begin{lstlisting}[basicstyle=\scriptsize\ttfamily]
foo:
 ...
 # Prepare arguments
 callq bar
 ...
  \end{lstlisting}
\end{lrbox}

\newsavebox{\modifiedcallbox}
\begin{lrbox}{\modifiedcallbox}
  \begin{lstlisting}[basicstyle=\scriptsize\ttfamily]
foo_0:
 ...
 # Prepare arguments
 callq *.Lptr_array+16(%rip)
 ...
  \end{lstlisting}
\end{lrbox}

\newsavebox{\controlflowtrampbox}
\begin{lrbox}{\controlflowtrampbox}
  \begin{lstlisting}[showlines=true,basicstyle=\scriptsize\ttfamily]
bar:
 # Preserve arguments
 jmpq *.Lptr_array+16(%rip)


\end{lstlisting}
\end{lrbox}

\begin{figure}
  \centering
  \subfloat[Original call site]{\usebox{\origcallbox}}
  \hfill
  \subfloat[Transformed call site]{\usebox{\modifiedcallbox}}
  \hfill
  \subfloat[Control-flow trampoline]{\usebox{\controlflowtrampbox}}
  \caption{Generated \arch~assembly}\label{fig:machine_code}
\end{figure}

\paragraph{\textbf{Embedding Metadata}}
Our runtime component needs to know which function variants are associated with each slot of the variant pointer array.
Depending on the partitioning policy, it may also require function execution frequencies and estimated execution costs for all function variants.
We add this information (encoded in an array of function descriptors) as read-only data to each compilation unit.

\subsection{The \toolname{} Runtime}

The \toolname{} runtime implements the three partitioning policies described in Section~\ref{sec:policy}.
The runtime exposes a single externally visible function used to register modules:
\code{cf\_register(const func\_t* start, const func\_t* end)}. 
Every module registers its function variants with the runtime by invoking this function from a constructor.
After all modules have registered, the runtime initializes.

The runtime's initialization proceeds in four steps.
First, the runtime computes the activation probabilities for each function variant, according to the configured policy.
Then, we seed a secure number generator.
Next, we initialize all variant pointer arrays.
This is necessary because the program might call some of the variant functions before our runtime's background thread performs its first round of run-time partitioning.
Finally, we spawn the background thread that is responsible for the continuous run-time partitioning.

\paragraph{\textbf{Run-time Partitioning}}
Our background thread runs an infinite loop, which invokes the partitioning procedure whenever it wakes up.
This procedure iterates through the function descriptors for every registered module.
For every function, we generate a random integer number $X$ between $0$ and $100$, and use this to select one of the variants.
If the activation probabilities for the sanitized and unsanitized variants of a function are $0.01$ and $0.99$, respectively, then we will activate the sanitized variant if $X$ is less than $2$, and we will activate the unsanitized variant for values greather than $1$.
We write the pointer to the activated variant in the variant pointer array.

We attempt to reduce cache contention by performing the write only if necessary (i.e., only if the old and new value differ).
This adds a read dependency on the old pointer value which may slow down the background thread.
However, the execution of the background thread is not performance critical since it runs fully asynchronously with respect to the application threads. 

\section{Effectiveness}\label{sec:effectiveness}

We evaluate the effectiveness of \toolname\ with an empirical investigation of five CVEs~\cite{CVEDatabase}, including the infamous Heartbleed bug.
Table~\ref{tab:cves} shows the CVEs we tested.
Each of them was found in a popular real-world program and the types of vulnerabilities include stack-based overflows and information leaks on the heap.
We used \toolname{} to compile two versions of each program, applying \asan\ to the sanitized variants in one version and \ubsan\ in the other version, and we configured our runtime to enforce its expected-cost partitioning policy.
We detected four out of five vulnerabilities in the \asan\ version, and three out of five in the \ubsan\ version.
We then compiled a third version of the program with the same partitioning policy and applied both sanitizers to the sanitized variants.
This third version reliably detects three out of five CVEs.
The remaining two CVEs are detected in \perc{72} and \perc{6} of our test runs.

\begin{table*}
  \scriptsize
  \caption{Evaluated CVEs}\label{tab:cves}
  \begin{tabular}{llllr}
    \toprule
	    CVE \#          & Program\enspace(Submodule)                  & Vulnerability                     & Sanitizer	          & Detection     \\
    \midrule
      2016\=/6297       & Php 7.0.3\enspace(Zip extension)          & Integer ovf. $\rightarrow$ Stack ovf.      &  \ubsan, \asan\     &  \tperc{71.8} \\
      2016\=/6289       & Php 7.0.3\enspace(Core engine)            & Integer ovf. $\rightarrow$ Stack ovf.      &  \ubsan, \asan\     &  Always       \\
      2016\=/3191       & Php 7.0.3\enspace(Pcre extension)         & Stack overflow                    &  \asan\             &  \tperc{6.2}  \\
      2014\=/0160       & OpenSSL\,1.0.1f\enspace(Heartbeat ext.)   & Heap over-read                    &  \asan\             &  Always       \\
      2014\=/7185       & Python 2.7.7\enspace(Core library)        & Integer ovf. $\rightarrow$ Heap over-read  &  \ubsan\            &  Always       \\
    \bottomrule
  \end{tabular}
\end{table*}

\noindent For each of the selected CVEs we perform the following steps:
\begin{enumerate}
  \item\label{step:verify_cve}
    Verify vulnerability exposure
  \item\label{step:verify_cve_detection}
    Verify vulnerability detection
  \item\label{step:collect_profile_data}
    Collect profile data
  \item\label{step:verify_prob_detection}
    Evaluate vulnerability detection with \toolname\
\end{enumerate}
Each of the above steps requires a program version with different instrumentation.
In step~\ref{step:verify_cve}, we compile the vulnerable program without any instrumentation and verify that the vulnerability can be triggered.
To do this, we use the proof-of-concept scripts referenced in the CVE details. 

In step~\ref{step:verify_cve_detection}, we compile the program with \asan\ or \ubsan\ enabled, but without \toolname.
We run our test script from step~\ref{step:verify_cve} to verify that the vulnerability is detected by the sanitizer.

Our expected-cost partitioning policy greatly benefits from profile data,
so in step~\ref{step:collect_profile_data}, we use LLVM's built-in profiling facilities to create an instrumented version of the program for collecting profile data.
We use the tests that come with the program as the profiling workload.
For vulnerabilities in submodules/extensions, we only run the tests of the submodule to increase the chance of the vulnerable code being classified as hot
(since vulnerabilities in cold code are guaranteed to be detected).
The test suite of the vulnerable OpenSSL version does not cover the Heartbeat extension.
Therefore, if we run the the test suite as-is, the function that contains the Heartbleed vulnerability is never executed.
\toolname{} would therefore classify this function as cold and always sanitize it, which guarantees detection.
To be more conservative, we executed the vulnerable function $300$ times with benign input alongside the official test suite.

Next, in step~\ref{step:verify_prob_detection}, we compile the program with the sanitizer enabled under \toolname.
We use \toolname{}'s default configuration
to compile each of the programs.
This means that the program contains two variants of all functions, except those that are cold and those without memory accesses.
We only created sanitized variants for cold functions, and unsanitized variants for functions without memory accesses.
Finally, we execute our test script from step~\ref{step:verify_cve} a thousand times to measure the detection rate.

Out of the five vulnerabilities, \asan\ and \ubsan\ detect four and three respectively.
The three vulnerabilities detected by \ubsan\ all involve an integer overflow.
The overflown value usually represents the length of some buffer, which results in out-of-bounds buffer accesses.
The other two vulnerabilities are caused by a lack of bounds checking.
Note that although the last CVE is classified as a heap over-read, \asan\ does not detect it.
The reason is that the Python interpreter uses a custom memory allocator.
It requests large chunks of memory from the operating system and maintains its own free lists to serve individual requests.
Unfortunately, \asan\ treats each chunk as a single allocation and therefore is unable to detect overflows within a chunk.
This shows that there is value in using multiple sanitizers that can detect different causes of vulnerabilities.

Lastly, we want to note that three out of five vulnerabilities are in code that \toolname\ classifies as cold.
For those cases, we manually verified that \toolname{} only created the sanitized variant for the vulnerable functions.
Hence, those vulnerabilities are always reported.
This result supports \toolname's underlying assumption that most bugs hide in infrequently executed code.
In summary, our results show that we always detect bugs in cold code while bugs in hot code are detected probabilistically.
We argue that this is a valuable property in our envisioned usage scenario:
finding bugs in beta software during real usage with an acceptable performance overhead.
Note that probabilistic detection is a property afforded by dynamic, but not by static partitioning.

\section{Efficiency}\label{sec:perf_eval}
We evaluated the performance of \toolname-enabled programs using the \speclong{} integer benchmark suite~\cite{SPEC}.
Since \toolname\ clones code we also measured the size of the resulting binaries.
Memory overheads---%
a small constant amount for the background thread and a few bytes of metadata for every function---%
are negligible (less than \perc{1}) for all \spec~programs, so we do not report them.

We conducted all experiments on a host with an Intel Xeon E5\=/2660~CPU and 64\,GB of RAM running 64\=/bit Ubuntu~14.04.
We applied \asan\ and \ubsan\ to all of the benchmark programs.
We configure \ubsan\ to disable error recovery, which always aborts the program instead of printing a warning message and attempting to recover for a subset of failed checks.
For configurations including \ubsan\ we also configure \toolname{} to create variants of all functions, even those that do not access memory.
We use the expected-cost partitioning policy with
a sanitization budget of \perc{1}, which our runtime evenly divides across all functions.

To collect profile data we use LLVM's built-in profiling facilities on the \emph{training} workload of \spec.
Since our chosen partitioning policy greatly benefits from profile data, we make the same data available to the baseline configuration to make the comparison fair.
We compile all configurations, including the baseline, with profile-guided optimization enabled, supplying the same profile data for all configurations.
When measuring the runtime, we use the \emph{reference} workload, run each benchmark three times, and report the median.

\subsection{Performance}
Figure~\ref{fig:perf_plot_asan} and~\ref{fig:perf_plot_ubsan} show the run-time overheads for \asan\ and \ubsan\ with respect to the baseline for all \spec\ integer benchmarks.
The last column depicts the geometric mean over all benchmarks, which is additionally stated in percent by Table~\ref{tab:perf} for easier reference.

\begin{figure*}
  \centering

  \begin{tikzpicture}
  \pgfmathsetmacro{\ymax}{3.0}
\begin{axis}[
  ybar=1pt,
  bar width=4.8pt,
  width=\textwidth+1cm,
  yscale=0.5,
  enlarge x limits={abs=15pt},
  legend cell align=left,
  legend style={/tikz/every even column/.append style={column sep=0.5cm},
    at={(0.5,-0.6)},anchor=north,legend columns=4},
  symbolic x coords={left, perlbench,bzip2,gcc,mcf,gobmk,hmmer,sjeng,libquantum,h264ref,omnetpp,astar,xalancbmk,geomean, right},
  xtick=data,
  x tick label style={rotate=35,anchor=east,yshift=-6pt,xshift=3pt},
  y tick label style={
        /pgf/number format/.cd,
            fixed,
            fixed zerofill,
            precision=1,
        /tikz/.cd
  },
  ymax=\ymax,
  ymin=0.80,
    visualization depends on={rawy \as \rawy},
    nodes near coords={\pgfmathprintnumber\rawy},
    restrict y to domain*={
        \pgfkeysvalueof{/pgfplots/ymin}:\pgfkeysvalueof{/pgfplots/ymax}
    },
    nodes near coords greater equal only/.style={
        /pgf/number format/.cd,
          fixed,
          fixed zerofill,
          precision=1,
        /tikz/.cd,
        small value/.style={
            /tikz/coordinate,
        },
        every node near coord/.append style={
            check for small values/.code={
                \begingroup
                \pgfkeys{/pgf/fpu}
                \pgfmathparse{\pgfplotspointmeta<#1}
                \global\let\result=\pgfmathresult
                \endgroup
                %
                %
                \pgfmathfloatcreate{1}{1.0}{0}
                \let\ONE=\pgfmathresult
                \ifx\result\ONE
                    \pgfkeysalso{/pgfplots/small value}
                \fi
            },
            check for small values,
        },
    },
    nodes near coords greater equal only=\ymax,
]

\addplot coordinates {
  (perlbench,   3.89)
  (bzip2,       1.71)
  (gcc,         2.29)
  (mcf,         1.61)
  (gobmk,       2.03)
  (hmmer,       1.87)
  (sjeng,       2.01)
  (libquantum,  1.45)
  (h264ref,     2.19)
  (omnetpp,     2.24)
  (astar,       1.54)
  (xalancbmk,   2.36)
  (geomean,     2.03)
};

\addplot coordinates {
  (perlbench,   2.91)
  (bzip2,       1.05)
  (gcc,         1.88)
  (mcf,         1.18)
  (gobmk,       1.12)
  (hmmer,       1.03)
  (sjeng,       1.03)
  (libquantum,  1.01)
  (h264ref,     1.33)
  (omnetpp,     1.76)
  (astar,       1.01)
  (xalancbmk,   1.63)
  (geomean,     1.33)
};

\addplot coordinates {
  (perlbench,   2.74)
  (bzip2,       1.02)
  (gcc,         1.75)
  (mcf,         1.01)
  (gobmk,       1.12)
  (hmmer,       1.03)
  (sjeng,       1.05)
  (libquantum,  1.02)
  (h264ref,     1.23)
  (omnetpp,     1.73)
  (astar,       0.99)
  (xalancbmk,   1.44)
  (geomean,     1.27)
};

\addplot[black,line legend,sharp plot,update limits=false, style=dashed]
  coordinates {(left, 1.0) (right, 1.0)
};

\legend{\asan, \toolname\,+\,\asan, \asan\ w/o checks}
\end{axis}
\end{tikzpicture}
  \caption{\spec\ run-time overheads for \asan}\label{fig:perf_plot_asan}

  \begin{tikzpicture}
  \pgfmathsetmacro{\ymax}{3.0}
\begin{axis}[
  ybar=1pt,
  bar width=4pt,
  width=\textwidth+1cm,
  yscale=0.5,
  enlarge x limits={abs=15pt},
  legend cell align=left,
  legend style={/tikz/every even column/.append style={column sep=0.25cm},
    at={(0.48,-0.6)},anchor=north,legend columns=4},
  symbolic x coords={left, perlbench,bzip2,gcc,mcf,gobmk,hmmer,sjeng,libquantum,h264ref,omnetpp,astar,xalancbmk,geomean, right},
  xtick=data,
  x tick label style={rotate=35,anchor=east,yshift=-6pt,xshift=3pt},
  y tick label style={
        /pgf/number format/.cd,
            fixed,
            fixed zerofill,
            precision=1,
        /tikz/.cd
  },
  ymax=\ymax,
  ymin=0.80,
    visualization depends on={rawy \as \rawy},
    nodes near coords={\pgfmathprintnumber\rawy},
    restrict y to domain*={
        \pgfkeysvalueof{/pgfplots/ymin}:\pgfkeysvalueof{/pgfplots/ymax}
    },
    nodes near coords greater equal only/.style={
        /pgf/number format/.cd,
          fixed,
          fixed zerofill,
          precision=1,
        /tikz/.cd,
        small value/.style={
            /tikz/coordinate,
        },
        every node near coord/.append style={
            check for small values/.code={
                \begingroup
                \pgfkeys{/pgf/fpu}
                \pgfmathparse{\pgfplotspointmeta<#1}
                \global\let\result=\pgfmathresult
                \endgroup
                %
                %
                \pgfmathfloatcreate{1}{1.0}{0}
                \let\ONE=\pgfmathresult
                \ifx\result\ONE
                    \pgfkeysalso{/pgfplots/small value}
                \fi
            },
            check for small values,
        },
    },
    nodes near coords greater equal only=\ymax,
]

\addplot coordinates {
  (perlbench,   1.36)
  (bzip2,       1.45)
  (gcc,         1.46)
  (mcf,         1.22)
  (gobmk,       1.51)
  (hmmer,       2.59)
  (sjeng,       1.56)
  (libquantum,  2.10)
  (h264ref,     1.68)
  (omnetpp,     1.37)
  (astar,       1.72)
  (xalancbmk,   1.52)
  (geomean,     1.59)
};

\addplot coordinates {
  (perlbench,   1.25)
  (bzip2,       1.04)
  (gcc,         1.07)
  (mcf,         0.98)
  (gobmk,       1.13)
  (hmmer,       1.09)
  (sjeng,       1.11)
  (libquantum,  1.04)
  (h264ref,     1.09)
  (omnetpp,     1.16)
  (astar,       1.25)
  (xalancbmk,   1.65)
  (geomean,     1.14)
};

\addplot coordinates {
  (perlbench,   4.20)
  (bzip2,       2.07)
  (gcc,         2.81)
  (mcf,         1.86)
  (gobmk,       2.78)
  (hmmer,       3.57)
  (sjeng,       2.91)
  (libquantum,  3.23)
  (h264ref,     3.08)
  (omnetpp,     2.84)
  (astar,       2.33)
  (xalancbmk,   4.19)
  (geomean,     2.91)
};

\addplot coordinates {
  (perlbench,   2.98)
  (bzip2,       1.20)
  (gcc,         1.92)
  (mcf,         1.10)
  (gobmk,       1.19)
  (hmmer,       1.13)
  (sjeng,       1.13)
  (libquantum,  1.08)
  (h264ref,     1.44)
  (omnetpp,     1.82)
  (astar,       1.22)
  (xalancbmk,   2.33)
  (geomean,     1.46)
};

\addplot[black,line legend,sharp plot,update limits=false, style=dashed]
  coordinates {(left, 1.0) (right, 1.0)
};

\legend{\ubsan, \toolname\,+\,\ubsan, \asan\,+\,\ubsan, \toolname\,+\,\asan\,+\,\ubsan}
\end{axis}
\end{tikzpicture}
  \caption{\spec\ run-time overheads for \ubsan\ and \asan+\ubsan}\label{fig:perf_plot_ubsan}
\end{figure*}

\begin{table}
  \centering
  \caption{\spec\ run-time overheads}\label{tab:perf}
  \begin{tabular}{lr}
    \toprule
      Configuration &  Overhead \\
    \midrule
      \toolname\                      &   \tperc{  2}   \\
      \asan\                          &   \tperc{103}   \\
      \asan\ w/o checks               &   \tperc{ 27}   \\
      \toolname\,+\,\asan\            &   \tperc{ 33}   \\ 
      \ubsan\                         &   \tperc{ 59}   \\
      \toolname\,+\,\ubsan\           &   \tperc{ 14}   \\ 
      \asan\,+\,\ubsan\               &   \tperc{191}   \\ 
      \toolname\,+\,\asan\,+\,\ubsan\ &   \tperc{ 46}   \\ 
    \bottomrule
  \end{tabular}
\end{table}

\toolname{}'s partitioning without any sanitization (with two identical variants) incurs a \perc{2} overhead on average, with a maximum of \perc{9} for \code{gobmk}.

For the fully-sanitized versions of \asan\ and \ubsan\ (absent \toolname{}) we measured an average overhead of \perc{103} and \perc{59} respectively.
Note that the overhead introduced by \asan\ can be as much as \perc{289} for \code{perlbench}.

We also created a modified version of \asan\ that does not execute any checks.
The remaining overhead can be attributed to the maintenance of metadata and other bookkeeping tasks.
This configuration represents a lower bound on the run time achievable by \toolname\ since bookkeeping needs to be done in all variants.
\toolname\ stays close to this lower bound for many benchmarks even when using the expected-cost policy in its default configuration.
For the \toolname-enabled versions of \asan\ and \ubsan\ we measured an average overhead of \perc{33} and \perc{14} respectively.
This corresponds to a reduction of overhead levels by more than two thirds (\perc{68} and \perc{76})
with respect to the fully-sanitized versions.
We also include a configuration that enables both \asan\ and \ubsan\ in Figure~\ref{fig:perf_plot_ubsan} to show that \toolname\ can handle multiple sanitizers as long as they are compatible with each other.

\subsection{Binary Size}

Table~\ref{tab:cve_binary_sizes} gives an overview of the impact that \toolname\ has on binary size for real-world programs.
We state binary sizes of the programs used in our effectiveness evaluation for \asan\ and \ubsan\ with and without \toolname\ and the size increase in percent.
We can navigate the size versus performance trade-off by adjusting our threshold for hot code and
argue that (using our policy) the maximum size increase is limited by a factor of two (i.e., when all code is classified as hot).

\begin{table}
  \caption{\toolname\ program sizes (in kilobytes)}\label{tab:cve_binary_sizes}
  \centering
  \begin{tabular}{l@{\hskip9pt} r@{\hskip2pt}r@{\hskip5pt}r @{\hskip9pt} r@{\hskip2pt}r@{\hskip5pt}r} 
    \toprule
      Program             &  \multicolumn{3}{c}{\asan}                   & \multicolumn{3}{c}{\ubsan}                      \\
    \midrule
      Php 7.0.3           &  20,483 / &  21,983  &  (\tperc{\enspace7})  &   8,658 / &  12,536  &   (\tperc{45})   \\
      OpenSSL 1.0.1f      &  19,128 / &  25,579  &  (\tperc{       34})  &  12,153 / &  14,243  &   (\tperc{17})   \\
      Python 2.7.7        &  41,715 / &  54,717  &  (\tperc{       31})  &  22,033 / &  28,641  &   (\tperc{30})   \\
    \bottomrule
  \end{tabular}
\end{table}

The statically-linked \toolname\ runtime adds a constant overhead of $6$\,KB\@ to each binary.
Internally, our runtime depends on the pthread library to spawn the background partitioning thread.
Usually, this does not increase program size as libpthread is a shared library.

We also measured the size of the \spec\ benchmark binaries used in our performance evaluation.
Since the benchmarks are small programs, the increase in relative code size is dominated by the inclusion of the \asan/\ubsan\ runtimes.
Therefore the larger programs in the suite exhibit the highest increase
(\perc{9} for \code{gcc}/\asan, and
\perc{16} for \code{xalancbmk}/\ubsan).
The increase in binary size over all benchmarks (geometric mean) for \asan\ and \ubsan\ are \perc{2} and \perc{5} respectively.

\section{Use Case: Fuzzing}

Fuzzing is an important use case for sanitization.
A fuzzer repeatedly executes a program with random inputs in order to find bugs.
Inputs that exercise new code paths are stored in a corpus (coverage-guided), which is used to derive further inputs (evolutionary).
To aid bug detection, the program is usually compiled with sanitization.
The vast majority of individual fuzzing runs do not detect bugs or increase coverage, so fuzzers rely on executing lots of runs (i.e., throughput is important).
We applied \toolname\ to LLVM's~\libfuzzer\cite{libFuzzer}, an in-process, coverage-guided, evolutionary fuzzing engine, with the goal of improving fuzzing efficiency.

When we first applied \toolname{} to fuzzing we noticed that it represents a specific use case that benefits from a custom partitioning policy.
Specifically, the fuzzer requires the program to be executed with coverage instrumentation.
The gathered coverage data is similiar (but not equivalent) to the profile data used for our partitioning policy.
We adapted \toolname{} to use online coverage data instead of profile data, which has two advantages.
First, it simplifies the developer workflow since there is no need to collect profile data a~priori. 
Second, it allows us to continuously refine our partitioning decisions.
We integrated \toolname{} with \libfuzzer{} with minimal changes to the latter.
Additionally, the main fuzzing loop provides a natural place to make partitioning decisions.
We added a call into our runtime from the fuzzing loop, forgoing the background thread in favor of synchronous partitioning.

\subsection{Partitioning Policy}
Our policy for fuzzing is simple.
For most functions we generate three variants:
variant\,\circledb{1} with coverage instrumentation,
sanitized variant\,\circledb{2}, and
fast variant\,\circledb{3} without any instrumentation.
At startup we activate variant\,\circledb{1} for the whole program.
Whenever the fuzzer discovers an input that exercises new code, we temporarily activate variant\,\circledb{2} for all functions and re-execute the input.
Finally, if a function becomes fully-explored (i.e., all its basic blocks have been executed), we activate its variant\,\circledb{3}.

Our policy allows us to increase coverage efficiently compared to the original program whose functions contain both coverage and sanitization instrumentation.
As coverage increases, functions transition from variant\,\circledb{1} to \circledb{3}, speeding up execution of the well-explored parts of the program.
The downside of this approach is that it potentially reduces the chance of bug detection as well as coverage feedback to the fuzzer.
Consider an input that exposes a non-crashing bug without increasing coverage.
Under our policy, such inputs execute without sanitization.
Additionally, a function that we deem ``fully-explored'' might still provide useful coverage feedback to the fuzzer.
The reason is that \libfuzzer's coverage model is fine-grained (e.g., it includes execution counts)
while our notion of fully-explored is binary.

\subsection{Evaluation}
\begin{figure}
  \centering
  \includegraphics[width=\textwidth]{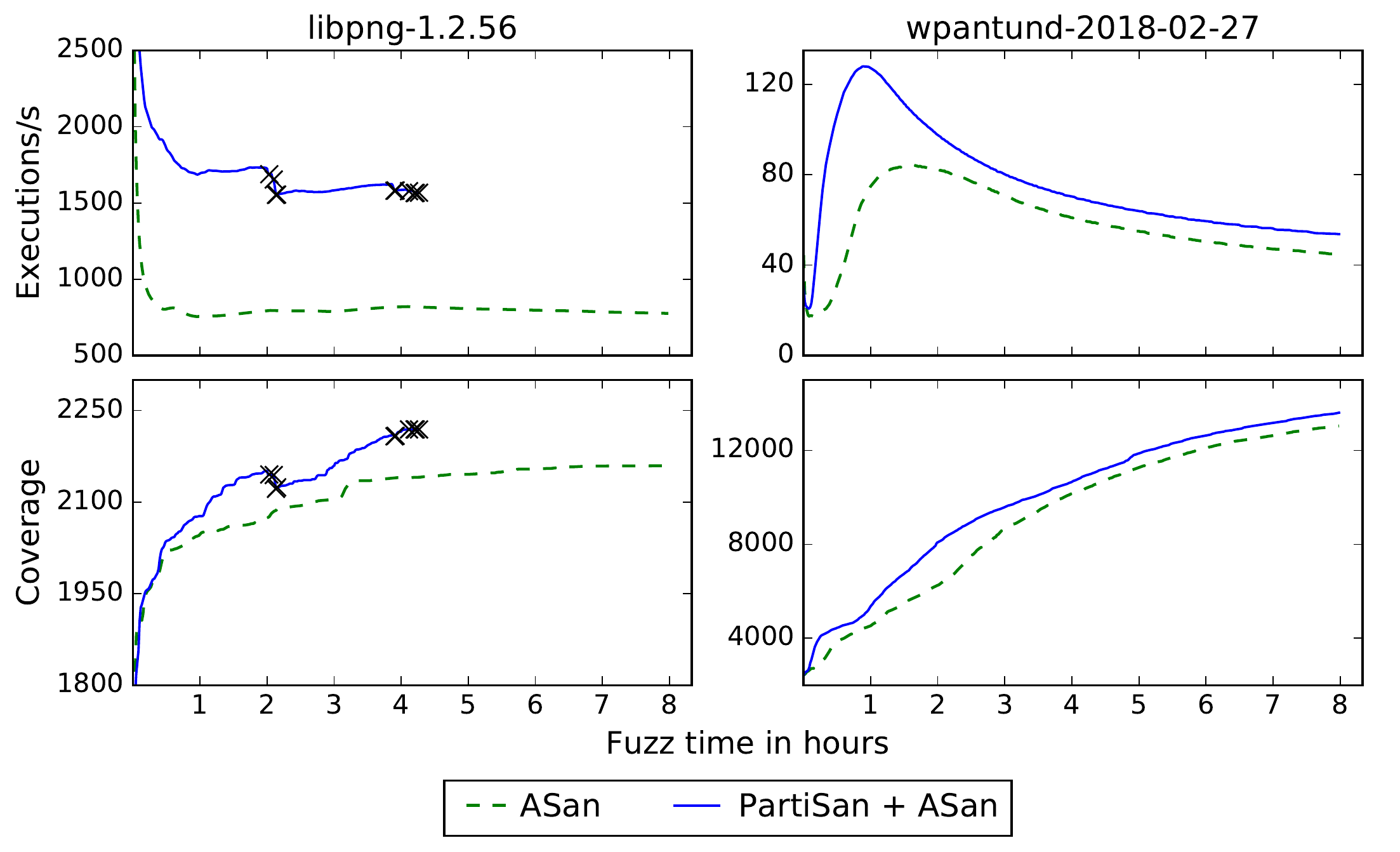}
  \caption{Fuzzing throughput and coverage for \code{libpng} and \code{wpantund}}\label{fig:fuzz_graphs}
\end{figure}

We evaluated the \toolname-enabled \libfuzzer{} on a popular benchmark suite for fuzzers~\cite{FTS} derived from widely-used libraries.
We ran all 23~included benchmarks with \asan{} enabled. 
Out of these 23~benchmarks 11~complete (find a bug) within a few minutes.
For the remaining 12~benchmarks we measured fuzzing throughput and coverage and ran them for eight hours or until completion.
Figure~\ref{fig:fuzz_graphs} shows the results for two benchmarks (geometric mean of 10~runs).
The markers indicate the completion of a run (i.e.,~after the first marker the line represents the remaining 9~runs).

As expected, \toolname{} is able to increase fuzzing throughput (executions per second) for the sanitized libraries.
For~9~(of~12) benchmarks this translates to improved coverage, and 3~benchmarks complete significantly faster.
For example, for the \code{libpng} benchmark (left side of Fig.\,\ref{fig:fuzz_graphs})
\toolname{} lets us find the bug within our time budget, whereas previously we could not.
However, the impact of \toolname{} is not always that pronounced.
For the \code{wpantund} benchmark (right side of Fig.\,\ref{fig:fuzz_graphs}), coverage only improves slightly.
Note that fuzzing throughput generally decreases over time as the fuzzer explores longer and longer code paths.

\section{Discussion}

\paragraph{Custom Partitioning Policies}
We implemented three run-time partitioning policies in \toolname{}.
The flexibility of our design and implementation additionally allows developers to define their own policies.
To implement a custom partitioning policy, the developer can provide her own \code{load\_policy} and \code{activate\_variant} function when linking the final binary.
Our policy for the fuzzing use case is built atop this mechanism.

\paragraph{Asynchronous Partitioning}
We opted to offload our run-time partitioning procedure onto a background thread.
The advantage of this approach is that, since partitioning happens asynchronously relative to the rest of the application, our runtime component has little impact on the application's performance.
The disadvantage is that we cannot partition on a per-function call basis or depending on the calling context.
That said, in the fuzzing use case we partition synchronously as part of the main fuzzing loop.

\paragraph{Partitioning Granularity}
\toolname{} partitions the program run time at function-level granularity.
In particular, \toolname{} might execute the sanitized variant of a hot function containing a long-running loop.
Executing this sanitized variant can induce a noticeable slowdown as \toolname{} does not support control-flow transfers between variants within the same function.
Our design can be refined with finer-grained partitioning, though a significant engineering effort would be required to implement it.
Our fundamental conclusions would not change with an improved partitioning scheme.

\paragraph{Selective Sanitization}\label{sec:limitations-tsan}
Like ASAP, \toolname{} does not support sanitizers that do not function correctly if they are applied selectively.
Consider, for example, a multithreaded program compiled with ThreadSanitizer~\cite{serebryany2009threadsanitizer}.
If two functions in the program concurrently write to the same memory location without acquiring a lock, then ThreadSanitizer will detect a data race.
This would not be true in a \toolname{}-enabled version of the program if we executed the sanitized variant of one function and the unsanitized variant of the other.
In this case, the data race would not be detected, thus rendering ThreadSanitizer ineffective.
\section{Related Work}

\subsection{Run-Time Partitioning}

Kurmus and Zippel proposed to create a split kernel with a protected partition containing a hardened variant of each kernel function, and an unprotected partition containing non-hardened variants~\cite{kurmus2014tale}.
Whenever the kernel services a system call or an interrupt request, it transfers control flow to one of the two partitions.
The protected (unprotected) partition is used to service requests from untrusted (trusted) processes and devices.
Unlike \toolname{}, however, it does not permit control flow transfers between the two partitions.
A service request is handled in its entirety by one of the two partitions.

The ASAP framework, presented by Wagner et al., reduces sanitizer overhead by removing sanitizer checks and programmer asserts from frequently executed code, while leaving the infrequently executed code unaffected~\cite{ASAP}.
This is also a form of partitioning, as ASAP creates a sanitized and an unsanitized partition within the program.
As with \toolname{}, transfers between sanitized and unsanitized code are frequent with ASAP\@.
However, contrary to \toolname{} and the aforementioned work, ASAP never creates multiple variants of a function.
ASAP should therefore be considered a static form of partitioning.
Note that static partitioning mechanisms can neither support adaptive overhead thresholds, nor probabilistic bug detection, nor our presented fuzzing policy.

Bunshin reduces sanitizer and exploit mitigation overhead by distributing security checks over multiple program variants and running them in parallel in an N-Variant execution system~\cite{xu2017bunshin}.
The key idea is to generate program variants in such a way that any specific sanitizer check appears in only one of the variants.
This distribution principle makes each variant faster than the original program and also enables the simultaneous use of incompatible tools.
Bunshin achieves full sanitizer coverage by running all variants in parallel, i.e., for any given sanitizer check there will be a variant that executes it.
This approach improves program latency at the cost of increased resource consumption which limits Bunshin's applicability.
In a fuzzing scenario, for example, available cores can be more efficiently leveraged by running additional fuzzer instances.

\subsection{Sanitizers}

We applied \toolname{} to two of the sanitizers that are part of the LLVM compiler framework, AddressSanitizer and Undefined\-Behavior\-Sanitizer~\cite{ASAN,UBSAN}.
Many other sanitizers exist.
MemorySanitizer detects reads of uninitialized values and, although we did not include it in our evaluation, it is fully compatible with \toolname{}~\cite{stepanov2015memorysanitizer}.
Sanitizers that detect bad casting~\cite{lee2015type, haller2016typesan, jeon2017hextype} and variadic function misuses~\cite{biswas2017venerable} could also benefit from \toolname{} by applying checks selectively.

ThreadSanitizer instruments memory accesses and atomic operations to detect data races, deadlocks, and misuses of thread synchronization primitives (e.g.,~pthread mutexes) in multithreaded programs~\cite{serebryany2009threadsanitizer}.
Unfortunately, it is not a good fit for \toolname{} because selective sanitization renders the sanitizer ineffective (cf. Section~\ref{sec:limitations-tsan}).

\subsection{Control-Flow Diversity}

\toolname{} partitions the run time of the protected program using control-flow diversity.
Prior work has explored the use of control-flow diversity for security purposes.
One such work, Isomeron~\cite{Isomeron}, is a defensive technique that defeats just-in-time return-oriented-programming (JIT-ROP) attacks~\cite{JIT-ROP}.
Isomeron creates diversified clones of the program's functions and switches randomly between functions on every function call and return statement.
Even with precise knowledge of the gadget locations, an attacker cannot mount a reliable JIT-ROP attack, as Isomeron might transfer control flow to a non-intended location after every execution of a gadget.

Crane et al.\ describe how they used control-flow diversity to mitigate cache-based side-channel attacks~\cite{StephenCacheSideChannels}.
Crane et al.\ create multiple variants of program functions and applies different diversifying transformations to each variant.
The transformations are designed to preserve the semantics of the code, but obscure the code's memory access patterns (i.e., data access locations and execution trace).
Essentially, the technique adds noise to the observable leakage in the shared cache, which raises the difficulty for the adversary.

\section{Conclusion}

We present \toolname, a run-time partitioning technique that increases the performance and flexibility of sanitized programs.
\toolname\ allows developers to ship a
single sanitizer-enabled binary \emph{without} having to commit to either the fraction of
time spent sanitizing on a given target, nor the type of sanitization
employed.
Specifically, \toolname\ uses run-time partitioning controlled by tunable policies. We have explored three simple policies and expect future developers to define additional, application and domain-specific ones.
Our experiments show that, using our expected-cost policy, \toolname\ reduces performance overheads of the two popular sanitizers, \asan\ and \ubsan, by \perc{68} and \perc{76} respectively.
We also demonstrate how \toolname\ can improve fuzzing efficiency.
When integrated with \libfuzzer, \toolname\ consistently increases fuzzing throughput which leads to improved coverage and more bugs found.

\toolname{}'s dynamic partitioning mechanism supports adaptive overhead thresholds and probabilistic bug detection;
neither of which are supported by static partitioning mechanisms presented in previous work.
Hence, \toolname\ is able to extend the usage scenarios of sanitizers to a much wider group of testers and their respective program inputs, leading to the exploration of a greater number of program paths.
This will enable developers to catch more errors early, reducing the number of vulnerabilities in released software.

{\balance\
  \bibliographystyle{splncs04}
\bibliography{indisan}}

\end{document}